\documentclass[journal]{IEEEtran}
\usepackage{cite}
\usepackage{amsmath,amssymb,amsfonts}
\usepackage{algorithmic}
\usepackage{graphicx}
\usepackage{textcomp}

\usepackage{times}
\usepackage{cite}
\usepackage{amssymb}
\usepackage{amsmath,epsfig}
\usepackage{subcaption}
\usepackage{float}
\usepackage{siunitx}
\usepackage{array}
\usepackage{setspace}
\usepackage{indentfirst}
\usepackage{algorithm}
\usepackage{algorithmic}
\usepackage{bm}
\usepackage{caption}
\usepackage{graphicx,xcolor}
\usepackage{color}
\usepackage{stfloats}

\renewcommand{\vec}[1]{\mathbf{#1}}

\DeclareMathAlphabet{\mathpzc}{OT1}{pzc}{m}{it}
\makeatletter
\newcommand*{\rom}[1]{\expandafter\@slowromancap\romannumeral #1@}
\makeatother
\usepackage{float}
\usepackage{multicol}
\DeclareMathOperator*{\argmax}{arg\,max}
\DeclareMathOperator*{\argmin}{arg\,min}
\usepackage{hyperref}

\begin{document}

\thispagestyle{empty}
\onecolumn
\begin{Large}IEEE Copyright Notice:\end{Large}

\vspace{1cm}
\textcopyright \ 2020 IEEE. Personal use of this material is permitted. Permission from IEEE must be obtained for all other uses, in any current or future media, including reprinting/republishing this material for advertising or promotional purposes, creating new collective works, for resale or redistribution to servers or lists, or reuse of any copyrighted component of this work in other works.

\vspace{1cm}
This article has been accepted for publication in a future issue of IEEE Transactions on Medical Imaging, but has not been fully edited. Content may change prior to final publication. Citation information:
DOI 10.1109/TMI.2020.3009022, IEEE Transactions on Medical Imaging.
URL: \url{https://ieeexplore.ieee.org/document/9139307}

\newpage

\twocolumn
\setcounter{page}{1}

\title{Approximating the Ideal Observer for joint signal detection and localization tasks by use of supervised learning methods}
\author{Weimin Zhou, Hua Li, and Mark A. Anastasio
\thanks{Copyright (c) 2019 IEEE. Personal use of this material is permitted. However, permission to use this material for any other purposes must be obtained from the IEEE by sending a request to pubs-permissions@ieee.org.}
\thanks{This work was supported in part by the NIH under Awards EB020604, EB023045, NS102213, EB028652, CA233873 and CA223799, and in part by the NSF under Award  DMS1614305. (Corresponding author: Mark A. Anastasio)}
\thanks{Weimin Zhou is with the Department of 
Electrical and Systems Engineering, Washington University in St. Louis, St. Louis,
MO, 63130 USA (e-mail: wzhou24@wustl.edu).}
\thanks{Hua Li is with the Department of Bioengineering, University of Illinois at
Urbana-Champaign, Urbana, IL 61801 USA, and also with the Carle
Cancer Center, Carle Foundation Hospital, Urbana, IL 61801 USA
(e-mail: huali19@illinois.edu).}
\thanks{Mark A. Anastasio is with the Department of 
Bioengineering, University of Illinois at Urbana-Champaign, Urbana,
IL, 61801 USA (e-mail: maa@illinois.edu).}}

\maketitle

\begin{abstract}
Medical imaging systems are commonly assessed and optimized by use of objective measures of image quality (IQ). The Ideal Observer (IO) performance has been advocated to provide a figure-of-merit for use in assessing and optimizing imaging systems because the IO sets an upper performance limit among all observers. When joint signal detection and localization tasks are considered, the IO that employs a modified generalized likelihood ratio test maximizes observer performance as characterized by the localization receiver operating characteristic (LROC) curve. Computations of likelihood ratios are analytically intractable in the majority of cases. Therefore, sampling-based methods that employ Markov-Chain Monte Carlo (MCMC) techniques have been developed to approximate the likelihood ratios. However, the applications of MCMC methods have been limited to relatively simple object models. Supervised learning-based methods that employ convolutional neural networks have been recently developed to approximate the IO for binary signal detection tasks. In this paper, the ability of supervised learning-based methods to approximate the IO for joint signal detection and localization tasks is explored. Both background-known-exactly and background-known-statistically signal detection and localization tasks are considered. The considered object models include a lumpy object model and a clustered lumpy model, and the considered measurement noise models include Laplacian noise, Gaussian noise, and mixed Poisson-Gaussian noise. The LROC curves produced by the supervised learning-based method are compared to those produced by the MCMC approach or analytical computation when feasible. The potential utility of the proposed method for computing objective measures of IQ for optimizing imaging system performance is explored.
\end{abstract}


\vspace{0.7cm}
\section{Introduction}
\label{sec:introduction}
\IEEEPARstart{M}{edical} imaging systems that produce images for specific diagnostic tasks 
are commonly assessed and optimized by use of objective measures of image quality (IQ).
Objective measures of IQ
quantify the performance of an observer for specific tasks\cite{barrett2013foundations, kupinski2003ideal, park2007channelized, park2009efficient, shen2006using, zhou2019approximating,zhou2018learning,zhou2019learningHO,zhou2020markov}. 
When imaging systems and data-acquisition designs are optimized for binary signal detection tasks (e.g., detection of tumor or lesion), 
the observer performance can be summarized by the receiver operating characteristic (ROC) curve.
The Bayesian Ideal Observer (IO) maximizes the area under the ROC curve (AUC) 
and has been advocated for computing figures-of-merit (FOMs) for guiding imaging system optimization~\cite{barrett2013foundations,kupinski2003ideal,zhou2019approximating}.
In this way, the amount of task-specific information in the measurement data is maximized.
The IO computes the test statistic as any monotonic transformation of the likelihood ratio\cite{barrett2013foundations}. 
It can be employed to assess efficiency of other numerical observers and human observers{\cite{burgess1981efficiency, liu1995object}}.

Joint signal detection and localization (detection-localization) tasks are frequently considered in medical
 imaging \cite{starr1975visual, gifford1999comparison, gifford2005comparison, zhou2009collimator, wunderlich2016optimal}.
When such tasks are considered,
the localization receiver operating characteristic (LROC) curve can be employed
 to describe the observer performance
and the area under the LROC curve (ALROC) can be employed
 as a FOM to guide the optimization of imaging systems. 
The IO employs a modified generalized likelihood ratio test (MGLRT) and maximizes the ALROC~\cite{khurd2005decision}.
Except for some special cases, the test statistics involved cannot be computed analytically.
Markov-Chain Monte Carlo (MCMC) techniques\cite{kupinski2003ideal} have been developed to approximate likelihood ratios by constructing Markov chains 
that comprise samples drawn from posterior distributions.
However, practical issues such as the design of proposal densities from which Markov chains can be efficiently generated need to be addressed. 
Current applications of MCMC methods for approximating the IO have been limited to specific object models 
that include lumpy object model\cite{kupinski2003ideal}, a parametrized torso phantom\cite{he2008toward}, and a binary texture model\cite{abbey2008ideal}.

Supervised learning-based methods hold great promise for establishing numerical observers 
that can be employed to compute objective measures of IQ.
When optimizing imaging systems and data-acquisition designs,
computer-simulated data are often employed~\cite{kupinski2003ideal,barrett2013foundations}. 
In such cases, large amounts of labeled data can be simulated to train inference models to be employed as numerical observers~\cite{kupinski2001ideal, zhou2019approximating}.  
Artificial neural networks (ANNs) possess the ability to represent complicated functions and, accordingly,
they can be trained to establish numerical observers and approximate test statistics.
Supervised learning methods have been successfully employed to train ANNs for this purpose.
For example, Kupinski \emph{et al.} explored
the use of fully-connected neural networks (FCNNs) to approximate the test statistic of the IO acting on low-dimensional feature vectors for binary signal detection tasks\cite{kupinski2001ideal}.
More recently, Zhou \emph{et al.} developed a supervised learning-based method 
for computing the test statistics of IOs performing binary signal detection
tasks with 2D image data by use of convolutional neural networks (CNNs)~\cite{zhou2018learning, zhou2019approximating}.
In addition,  the ability of CNNs to approximate the IO for a background-known-exactly (BKE) signal detection-localization task has been explored\cite{zhou2019learning_lroc}.
{However,
there remains a need to explore methods for approximating the IO
for signal detection-localization tasks that account for background variability.}

In this work, a supervised learning-based method that employs CNNs to approximate the IO 
for signal detection-localization tasks is explored.
The proposed method represents a deep-learning-based implementation of the IO decision strategy proposed in the seminal theoretical work by Khurd and Gindi \cite{khurd2005decision}.
The considered signal detection-localization tasks involve various object models in combination with several realistic measurement noise models.
Numerical observer performance is assessed via the LROC analysis.
The results of the proposed supervised-learning methods are compared to those produced by MCMC methods or analytical computations when feasible.

The remainder of this paper is organized as follows. 
Salient aspects of joint signal detection-localization theory
 are reviewed in Section~\ref{sec:background}.
The use of supervised deep learning for approximating the IO for 
signal detection-localization tasks is described in Section~\ref{sec:method}.
The numerical studies and results are discussed in Sections~\ref{sec:studies} and~\ref{sec:result}, respectively. 
Finally, the article concludes with a discussion in Section~\ref{sec:discussion}.
 
\vspace{1.1cm}
\section{Background}
\label{sec:background}
A digital imaging system can be  described as a continuous-to-discrete (C-D) mapping:
\begin{equation}\label{eq:H}
\vec{g} = \mathcal{H}f(\vec{r}) + \vec{n},
\end{equation}
where the vector $\vec{g}\in \mathbb{R}^M$ denotes the measured image data, 
$f(\vec{r})$ is a compactly supported and bounded function of a spatial coordinate $\vec{r}\in \mathbb{R}^d$ that describes
the object being imaged, $\mathcal{H}$ is the imaging operator that maps $\mathbb{L}_2(\mathbb{R}^d)\rightarrow \mathbb{R}^M$, and $\vec{n}\in \mathbb{R}^M$ is the measurement noise.
The measured data $\vec{g}$ is random because the measurement noise $\vec{n}$ is random.
Object variability is known to limit observer performance \cite{park2007efficiency}.
 As such, the object 
 function $f(\vec{r})$ can be either deterministic or stochastic, depending on
the specification of the diagnostic task to be assessed. When linear imaging operators are considered, 
the imaging process described in Eqn. (\ref{eq:H}) can be written as \cite{barrett2013foundations}:
\begin{equation}
g_m = \int_{\mathbb{R}^d} d\vec{r}\ h_m(\vec{r}) f(\vec{r}) + n_m,
\end{equation}
where $g_m$ and $n_m$ are the $m^{th}$ element of $\vec{g}$ and $\vec{n}$, respectively, 
and {$h_m(\vec{r})$ is the sensitivity function and also called the point response function (PRF) that describes the sensitivity of the $m^{th}$ measurement to the function $f(\vec{r})$ at point $\vec{r}$.}
\subsection{Detection-localization tasks with a discrete-location model}
Detection-localization tasks 
in which the signal location is modeled as a discrete parameter having finite possible values are considered~\cite{khurd2005decision}.
Also, for the tasks considered, signal-present images are assumed
to contain a single signal~\cite{khurd2005decision}.
The number of possible signal locations is denoted as $J$. 
An observer is required to classify an image as satisfying either 
one of the $J+1$ hypotheses (i.e., one signal-absent hypothesis and $J$ signal-present hypotheses). 
The imaging processes under these hypotheses can be represented as:
\begin{equation}\label{eq:imgH}
\begin{split}
&H_{0}:  \mathbf{g} = \mathbf{b}+ \mathbf{n}, \\
&H_{j}:  \mathbf{g} = \mathbf{b}+\mathbf{s}_j + \mathbf{n}, 
\end{split}
\end{equation}
where $j = 1,2,...,J$, $\vec{b} \equiv \mathcal{H}f_b(\vec{r})$ is the image of the background
  $f_b(\vec{r})$, and $\vec{s}_j \equiv \mathcal{H}f_{s_j}(\vec{r})$ is the 
image of the signal $f_{s_j}(\vec{r})$ at the $j^{th}$ location. 
The quantities $b_m$ and $s_{j_m}$ will denote the $m^{th}$ element of $\vec{b}$ and $\vec{s}_j$, respectively.
It should be noted that $H_0$ is the signal-absent hypothesis, and $H_j$ is the signal-present hypothesis corresponding to the $j^{th}$ signal location ($j = 1,2,...,J$).

Scanning observers compute a test statistic for each possible signal location 
and a max-statistic rule can be subsequently implemented to make a decision~\cite{gifford2016visual}.
The decision strategy employed by scanning observers
with max-statistic rules is described as\cite{gifford2016visual}:
\begin{equation}
\begin{split}
&t(\mathbf{g}) =  \underset{j\in\{1,...,J\}}{\text{max} } \lambda_j(\vec{g})  \\
&j^*(\mathbf{g}) =\argmax_{j\in\{1,...,J\}} \lambda_j(\vec{g})\\
\text{Decide } &\text{$H_{j^*(\mathbf{g})}$ if $t(\mathbf{g}) > \tau$, else decide $H_0$}.
\end{split}
\end{equation}
Here, $\lambda_j(\vec{g})$ maps the measured image $\vec{g}$ to a real-valued test statistic corresponding to the $j^{th}$ signal location, and $\tau$ is a predetermined decision threshold.
{An LROC curve that depicts the tradeoff between the probability of correct localization and the false-positive rate is formed by varying $\tau$, and the ALROC can be subsequently computed as a FOM to quantify the observer performance \cite{khurd2005decision}. The ALROC can also be computed from a two-alternative forced-choice (2AFC) test without plotting the LROC curve \cite{clarkson2007estimation}.}

\subsection{Scanning Ideal Observer and Scanning Hotelling Observer}
The scanning IO that employs a modified generalized likelihood ratio test (MGLRT) maximizes the signal detection-localization task performance {as measured by the ALROC. }
The MGLRT\cite{khurd2005decision} can be described as:
\begin{equation}\label{eq:GLRT}
\begin{split}
&t_{{LR}}(\mathbf{g})  =  \max_{j\in\{1,...,J\}} \frac{\Pr(H_j)p(\mathbf{g}|H_j)}{p(\mathbf{g}|H_0)} \\
&j_{{LR}}^*(\mathbf{g}) =\argmax_{j\in\{1,...,J\}}\frac{\Pr(H_j)p(\mathbf{g}|H_j)}{p(\mathbf{g}|H_0)}\\
\text{Decide }&\text{$H_{j_{{LR}}^*(\mathbf{g})}$ if $t_{{LR}}(\mathbf{g}) > \tau_{{LR}}$, else decide $H_0$}.
\end{split}
\end{equation}
By use of Bayes rule, the decision strategy described in Eqn. (\ref{eq:GLRT}) is equivalent to a posterior ratio test\cite{zhou2019learning_lroc}:
\begin{equation}\label{eq:GLRT2}
\begin{split}
&t_{{PR}}(\mathbf{g}) =  \max_{j\in\{1,...,J\}} \frac{\Pr(H_j|\mathbf{g})}{\Pr(H_0|\mathbf{g})}  \\
&j_{{PR}}^*(\mathbf{g}) =\argmax_{j\in\{1,...,J\}}\frac{\Pr(H_j|\mathbf{g})}{\Pr(H_0|\mathbf{g})} \\
\text{Decide } &\text{$H_{j_{{PR}}^*(\mathbf{g})}$ if $t_{{PR}}(\mathbf{g}) > \tau_{{PR}}$, else decide $H_0$}.
\end{split}
\end{equation}
{The case $\tau_{PR} = 1$ minimizes the probability of error
and the corresponding decision rule is called minimum-error criterion or the maximum a posteriori probability (MAP) criterion\cite{barrett2013foundations}.}

When the likelihood function $p(\vec{g}|H_j)$ in Eqn. (\ref{eq:GLRT}) can be described by a Gaussian probability density function having the same covariance matrix $\bm{\vec{K}}$ under each hypothesis,
the IO is equivalent to a scanning Hotelling Observer (HO)~\cite{barrett2006objective, gifford2014efficient, gifford2016visual}. 
{Denote the mean of background $\vec{b}$ as $\bar{\vec{b}} = \langle \vec{b} \rangle_{\vec{b}}$, when $\Pr(H_j)$ is a constant, the scanning HO can be represented as~\cite{barrett2006objective, gifford2014efficient, gifford2016visual}:}
\begin{equation}\label{eq:sho}
\begin{split}
&{t_{HO}(\mathbf{g}) =  \max_{j\in\{1,...,J\}} \vec{w}_{HO_j}^T \left(\vec{g} - \bar{\vec{b}} - \frac{\vec{s}_j}{2}\right)}\\
&{j_{HO}^*(\mathbf{g}) =\argmax_{j\in\{1,...,J\}}\vec{w}_{HO_j}^T \left(\vec{g} - \bar{\vec{b}} - \frac{\vec{s}_j}{2}\right)}\\
\text{Decide }&{\text{$H_{j_{HO}^*(\mathbf{g})}$ if $t_{HO}(\mathbf{g}) > \tau_{HO}$, else decide $H_0$},}
\end{split}
\end{equation}
where $\vec{w}_{HO_j} \equiv \bm{\vec{K}}^{-1}\vec{s}_j$ is the Hotelling
 template corresponding to the $j^{th}$ signal location. 
Due to its relative ease of implementation, the scanning HO can be employed when the IO test statistic is difficult to compute. 
In addition, for a simplified binary signal detection task, 
{a possible IO test statistic can be computed as a posterior probability $\Pr(H_{present}|\vec{g})= \sum_{j=1}^{J}{\Pr(H_j|\mathbf{g})}\equiv 1 - \Pr(H_0|\mathbf{g})$, which is a monotonic transformation of the likelihood ratio.}

{It should be noted that the considered scanning observers described above 
correspond to a discrete-location model without search tolerance.
Detailed discussions on search tolerance can be found in the reference \cite{khurd2005decision}.
The considered observers can be generalized to location models that include search tolerance according to the reference\cite{khurd2005decision}.}

\vspace{0.3cm}
\section{Approximating the IO for signal detection-localization tasks by use of CNNs}
\label{sec:method}
To approximate the IO for signal detection-localization tasks,
a CNN can be trained to approximate a set of posterior probabilities that are employed in the posterior probability test  in Eqn. (\ref{eq:GLRT2}).
To achieve this, the softmax function is employed in the last layer of the CNN, the so-called softmax layer{~\cite{rawat2017deep}},
so that the output of the CNN can be interpreted as probabilities.
Let $\bm\Theta$ denote the vector of weight parameters of a CNN
 and let $\vec{z}(\vec{g}; \bm\Theta) \in \mathbb{R}^{J+1}$
 denote the output of the last hidden layer of the CNN, 
 which is also the input to the softmax layer.
The CNN-approximated posterior probabilities can be computed as: 
\begin{equation}
\Pr(H_j | \vec{g}, \bm\Theta) \equiv \frac{\exp[{z_j(\vec{g}; \bm\Theta)}]}{\sum_{j'=0}^{J}\exp[{z_{j'}(\vec{g}; \bm\Theta)}]},\ j = 0,1, ..., J, 
\label{eq:CNN-posterior}
\end{equation}
where $z_j(\vec{g}; \bm\Theta)$ is the $(j+1)^{th}$ element of $\vec{z}(\vec{g}; \bm\Theta)$.
The CNN parameter vector $\bm\Theta$ is to be determined such that the difference between the 
CNN-approximated posterior probability $\Pr(H_j | \vec{g}, \bm\Theta)$ and the actual posterior probability $\Pr(H_j | \vec{g})$ is minimized.

The maximum likelihood (ML) estimate of $\bm\Theta$ can be approximated by use of a supervised learning method{\cite{kupinski2001ideal}}.
Let $y\in \{0,1,...,J\}$ denote the label of the measured image $\vec{g}$, where $y = j$ corresponds to the hypothesis $H_j$. 
{Given the joint probability distribution $p(\vec{g}, H_y)$,} the ML estimate of the CNN weight parameters $\bm\Theta_{ML}$
can be obtained by minimizing the generalization error, which is defined as the ensemble average of the cross-entropy over the distribution $p(\vec{g}, {H_y})$\cite{kupinski2001ideal,zhou2019approximating}:
\begin{equation}\label{eq:ml}
\bm\Theta_{ML} = \argmin_{\bm\Theta} \langle-\log[\Pr(H_y|\vec{g}, \bm\Theta)]\rangle_{(\vec{g}, y)},
\end{equation}
where $\langle \cdot \rangle_{(\vec{g}, y)}$ denotes the ensemble average over the distribution $p(\vec{g}, {H_y})$. When the CNN possesses sufficient representation capacity such that $\vec{z}(\vec{g}; \bm\Theta)$ can take any functional form, 
$\Pr(H_j | \vec{g}, \bm\Theta_{ML}) = \Pr(H_j | \vec{g})$.
To see this, one can 
compute the gradient of the cross-entropy with respect to ${z}_j(\vec{g}; \bm\Theta)$ as:
\begin{equation}\label{eq:gradient}
\begin{aligned}
\frac{\partial  \langle-\log[\Pr({H_y}|\vec{g}, \bm\Theta)]\rangle_{(\vec{g}, y)}}{\partial z_j(\vec{g}; \bm\Theta)}&
\\
= p(\vec{g})\Big[\frac{\exp[{z_j(\vec{g}; \bm\Theta)}]}{\sum_{j'=0}^{J}\exp[{z_{j'}(\vec{g}; \bm\Theta)}]} &- \Pr(H_j|\vec{g})\Big].
\end{aligned}
\end{equation}
The derivation of this gradient computation can be found in Appendix \ref{append}.
Because $\vec{z}(\vec{g}; \bm\Theta)$ can take any functional form when the CNN possesses sufficient
 representation capacity,
determining $\bm\Theta_{ML}$ involves finding $\vec{z}(\vec{g}; \bm\Theta)$ that minimizes the cross-entropy defined in Eqn. (\ref{eq:ml}).
According to Eqn. (\ref{eq:gradient}), for any $\vec{g}\in \{\vec{g}| p(\vec{g})\neq 0 \}$, the optimal solution $z_j(\vec{g}; \bm\Theta_{ML})$ that has zero gradient value
satisfies $\frac{\exp[{z_j(\vec{g}; \bm\Theta_{ML})}]}{\sum_{j'=0}^{J}\exp[{z_{j'}(\vec{g}; \bm\Theta_{ML})}]}  = \Pr(H_j|\vec{g})$,
from which $\Pr(H_j | \vec{g}, \bm\Theta_{ML}) = \Pr(H_j|\vec{g})$.

Given a training dataset that contains $N$
 independent training samples $\{ (\vec{g}_i, y_i) \}_{i=1}^{N}$,
$\bm\Theta_{ML}$ can be estimated by minimizing the empirical error as:
\begin{equation}\label{eq:loss}
\hat{\bm\Theta}_{ML} = \argmin_{\bm\Theta} \frac{1}{N}\sum_{i=1}^{N}-\log[\Pr({H_{y_i}}|\vec{g}_i, \bm\Theta)],
\end{equation}  
where $\hat{\bm\Theta}_{ML}$ is an empirical estimate of $\bm\Theta_{ML}$.
The posterior probability $\Pr(H_j|\vec{g})$ can be subsequently approximated by the CNN-represented posterior probability $\Pr(H_j | \vec{g}, \hat{\bm\Theta}_{ML})$
and the decision strategy described in Eqn. (\ref{eq:GLRT2}) can be implemented.
It should be noted that minimizing empirical error on a small training dataset can result in overfitting and large generalization errors. 
Mini-batch stochastic gradient descent algorithms can be employed to reduce the rate of overfitting~\cite{goodfellow2016deep}. 
These mini-batches can be generated on-the-fly when online learning is implemented.

\vspace{0.8cm}
\section{Numerical studies}
\label{sec:studies}
Computer-simulation studies were conducted to investigate the supervised learning-based method for approximating the IO for 
signal detection-localization tasks. The considered signal detection-localization tasks
 included two background-known-exactly (BKE) tasks and two
background-known-statistically (BKS) tasks.
A lumpy background (LB) model \cite{kupinski2005small} and a clustered lumpy background (CLB) model \cite{bochud1999statistical} were employed in the BKS tasks. 

The imaging system considered was an idealized parallel-hole collimator system that
 was described  by a linear C-D mapping with  
Gaussian point response functions (PRFs) given by \cite{kupinski2003ideal, kupinski2003experimental}:
\begin{equation}
h_m(\vec{r}) = \frac{\mathpzc{h}}{2\pi w_h^2}\exp\left(-\frac{(\vec{r}-\vec{r}_m)^T(\vec{r}-\vec{r}_m)}{2w_h^2}  \right),
\end{equation}
where $\mathpzc{h}$ and $w_h$ are the height and width of the PRFs, respectively.
{Imaging systems with larger $\mathpzc{h}$ have greater sensitivity while imaging systems with larger $w_h$ have lower resolution.}

{Denote $\tilde{\vec{r}}_j=\mathbf{R}_{\theta_j} (\vec{r}-\vec{r}_{c_j})$,}
the signal to be detected and localized was modeled by a 2D Gaussian function with 9 possible locations:
\begin{equation}
\label{eq:signal}
{f_{s_j}(\vec{r}) = {a_{s_j}}\exp\left(- \tilde{\vec{r}}_j^T\mathbf{D}_j^{-1} \tilde{\vec{r}}_j \right),}
\end{equation}
where $a_{s_j}$ is the signal amplitude, 
$\vec{R}_{\theta_j}= \begin{bmatrix}
       \cos(\theta_j) & -\sin(\theta_j)           \\[0.3em]
       \sin(\theta_j) & \cos(\theta_j)
     \end{bmatrix}$
is {a rotation matrix} corresponding to the rotating angle $\theta_j$,
$\vec{D}_j=
\begin{bmatrix}
2w_{1_j}^2 & 0 \\
0 & 2w_{2_j}^2 
\end{bmatrix}
$  is a matrix that determines the width of the $j^{th}$ signal along each axis,
and $\vec{r}_{c_j}$ is the center location of the $j^{th}$ signal.
With consideration of the specified imaging system, the $m^{th}$ element ${s}_{j_m}$ of the signal image ${\vec{s}_j}$ can be subsequently computed as:
 \begin{equation}
{ {s_j}_m  =  A_j  \exp{\left({-\tilde{\vec{r}}_{jm}^{T}\vec{D}_j'^{-1}\tilde{\vec{r}}_{jm}}\right)},}
 \end{equation} 
where $A_j = {a_{s_j}\mathpzc{h}w_{1_j} w_{2_j}} \sqrt{\frac{1}{(w_h^2+w_{1_j}^2)(w_h^2+w_{2_j}^2)}}$, $\vec{D}_j'= \begin{bmatrix}
2(w_h^2+w_{1_j}^2) & 0 \\
0 & 2(w_h^2+w_{2_j}^2) 
\end{bmatrix}$
and {$\tilde{\vec{r}}_{jm}=\mathbf{R}_{\theta_j} (\vec{r}_m-\vec{r}_{c_j})$.}

For each task described below,  the LROC curves were fit by use of LROC software \cite{Judy2010LROC} that implements Swensson's fitting algorithm\cite{swensson1996unified} and the IO performance was quantified by the
ALROC. 

\vspace{0.4cm}
\subsection{BKE signal detection-localization tasks}
\begin{figure}[ht]
\centering
\begin{subfigure}[b]{0.24\textwidth}
   \centering
 \includegraphics[width=1.0\linewidth]{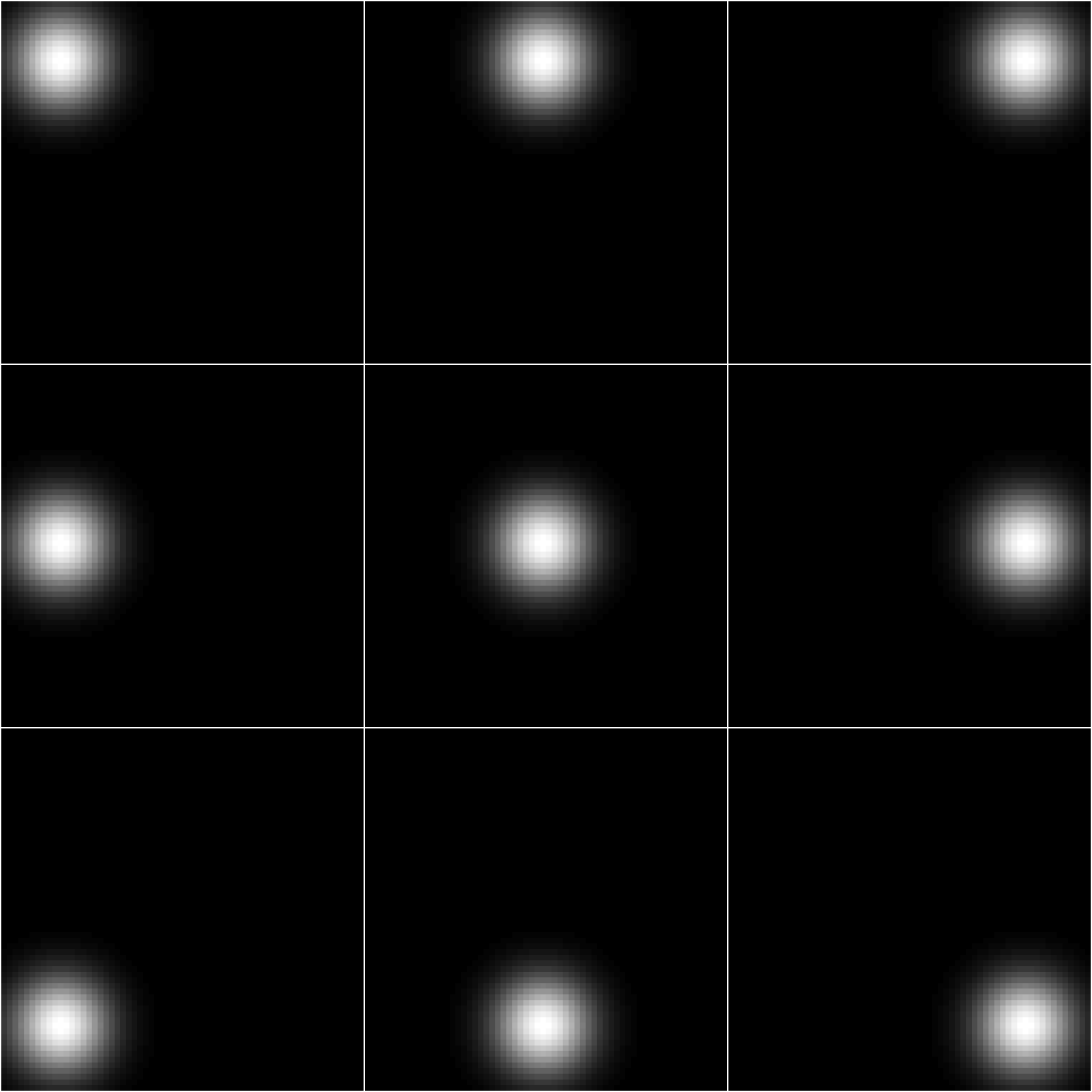}
 \caption{}
 \end{subfigure}
 \begin{subfigure}[b]{0.24\textwidth}
  \centering
 \includegraphics[width=1.0\linewidth]{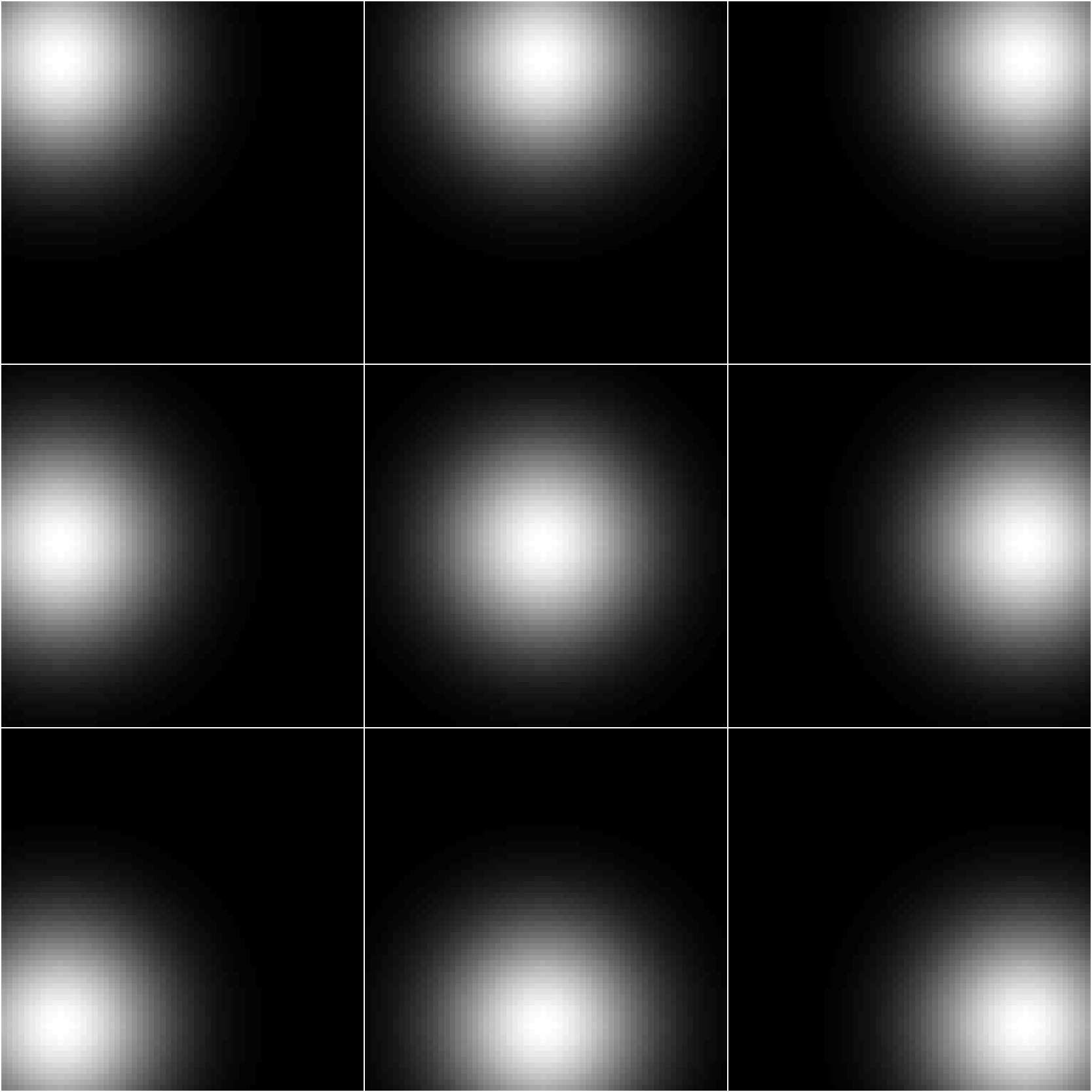}
  \caption{}
 \end{subfigure}
 \caption{(a) Signal images corresponding to ``System 1".  (b) Signal images corresponding to ``System 2"}
 \label{sig_imgs}
\end{figure}

\vspace{0.05cm}
For the BKE tasks, the size of background image was $64\times 64$ pixels and $\vec{b} = \vec{0}$.
The signal to be detected and localized had the signal amplitude $a_{s_j} = 0.2$, width $w_{1_j} = w_{2_j} = 3$, and $\vec{R}_{\theta_j} = 0$ for all 9 possible locations $j=1,2,...,9$.
Two imaging systems described by different PRFs  were considered.
The first imaging system, ``System 1'', was described by $\mathpzc{h}=60$ and $w_h=5$.
The second imaging system, ``System 2", was described by $\mathpzc{h}=144$ and $w_h=12$.
The signals at different locations imaged through the two imaging systems are illustrated in Fig. \ref{sig_imgs}.

To investigate the ability of the CNN to approximate a non-linear IO test statistic,
a Laplacian probability density function,
which has been utilized  
to describe histograms of 
fine details in 
digital mammographic images \cite{heine1999multiresolution, clarkson2000approximations}, 
was employed to model the likelihood function $p(\vec{g}|H_j)$.
Specifically, the measured image data $\vec{g}$ were simulated by adding independent and identically distributed (i.i.d.) Laplacian noise\cite{clarkson2000approximations}:
$n_m\sim L(0,c)$, where $L(0,c)$ denotes a Laplacian distribution with the mean of 0 and the exponential decay of c, which was set to $20/\sqrt{2}$ corresponding to a standard deviation of 20.
In this case, the likelihood ratio can be analytically computed as\cite{clarkson2000approximations}:
\begin{equation}\label{eq:bke}
\Lambda_j(\vec{g}) = \exp\left[\frac{1}{c}\sum_{m=1}^{M} (|g_m - b_m| - |g_m-b_m-s_{j_m}|) \right].
\end{equation}
The IO decision strategy described by Eqn. (\ref{eq:GLRT}) was subsequently implemented by use of the likelihood ratios given by Eqn. (\ref{eq:bke}), and the resulting LROC curves and ALROC
values were compared to those produced by the proposed supervised learning method described in Sec. \ref{sec:method}.

{Rank ordering of imaging system designs depends on specifications of tasks\cite{myers1990aperture}.}
The two imaging systems were ranked by use of the IO performance for the considered signal detection-location tasks via LROC analysis.
In addition, 
to demonstrate that the imaging system design
optimized by use of the IO for signal detection-localization tasks may differ from
that optimized by use of the IO for the simplified binary signal detection tasks,
the two imaging systems were also assessed by use of the IO performance
for the simplified binary signal detection tasks via ROC analysis.
The ROC curves were fit by use of Metz-ROC software\cite{metz1998rockit} using ``proper" binormal model\cite{metz1999proper}.

\vspace{0.8cm}
\subsection{BKS signal detection-localization task with a lumpy background model}
The first  BKS task utilized a lumpy object model to emulate background variability \cite{kupinski2005small}. 
The considered lumpy background models are described as\cite{barrett2013foundations,kupinski2005small}:
\begin{equation}
f_b(\vec{r}) = \sum_{n=1}^{N_b}l(\vec{r}-\vec{r}_n|a, w_b),
\end{equation} 
where $N_b\sim P(\bar{N})$ denotes the number of the lumps, $P(\bar{N})$ denotes a Poisson distribution with mean of $\bar{N} = 8$, and $l(\vec{r}-\vec{r}_n|a, w_b)$
denotes the lump function that was modeled by a 2D Gaussian function:
\begin{equation}
l(\vec{r}-\vec{r}_n|a, w_b) = {a}\exp\left(-\frac{(\vec{r}-\vec{r}_n)^T(\vec{r}-\vec{r}_n)}{2w_b^2}  \right).
\end{equation}
Here, $a = 1$, $w_b = 7$, and $\vec{r}_n$ denotes the center location of the $n^{th}$ lump that was sampled from a uniform distribution over the image field of view.
The imaging system  PRF was specified  by $\mathpzc{h} = 40$ and  $w_h = 1.5$. 
The image size was $64\times 64$ and 
 the $m^{th}$ $(1\leq m \leq 4096)$ element of the background image ${b}_m$ is given by:
\begin{equation}
b_{m} = \frac{a\mathpzc{h}w_b^2}{w_h^2+w_b^2}\exp\left(-\frac{(\vec{r}_m-\vec{r}_n)^T(\vec{r}_m-\vec{r}_n)}{2(w_h^2+w_b^2)}  \right).
\end{equation}

The measurement noise considered in this case was i.i.d. Gaussian noise with a mean of 0 and a standard deviation of 20. 
Three realizations of the signal-absent images are shown in the top row in Fig. \ref{fig:imgs}.
The signals to be detected and localized were specified by Eqn.\ (\ref{eq:signal})
 with $a_{s_j}=0.5$, $w_{1_j} = w_{2_j} = 2$, and $\vec{R}_{\theta_j} = 0$ for all 9 possible locations $j=1,2,...,9$.  The signal at different locations is illustrated in Fig. \ref{fig:CLB_sigs} (a).

Because the likelihood ratios $\Lambda_j(\vec{g})$ in this case cannot be analytically computed,
 the MCMC method developed by Kupinski \emph{et al.}\cite{kupinski2003ideal} was
 implemented  as a reference method.
The MCMC method computed the likelihood ratio as:
\begin{equation}
\Lambda_j(\vec{g}) \approx \frac{1}{N_c}\sum_{i=1}^{N_c} \Lambda_{\text{BKE}_j}(\vec{g}|\vec{b}^{(i)}),
\end{equation}
where $\Lambda_{\text{BKE}_j}(\vec{g}|\vec{b}^{(i)})\equiv \frac{p(\vec{g}|\vec{b}^{(i)}, H_j)}{p(\vec{g}|\vec{b}^{(i)}, H_0)}$ is the BKE likelihood ratio conditional on the $i^{th}$ background image $\vec{b}^{(i)}$
and $N_c$ is the number of samples used in Monte Carlo integration.
Because Gaussian noise was considered in this case, $\Lambda_{\text{BKE}_j}(\vec{g}|\vec{b}^{(i)})$ can be analytically computed as:
\begin{equation}
\Lambda_{\text{BKE}_j}(\vec{g}|\vec{b}^{(i)}) = \exp \left[(\vec{g}-\vec{b}^{(i)}-\vec{s}_j/2)^T\mathbf{K}_n^{-1}\vec{s}_j \right],
\end{equation}
where $\mathbf{K}_n$ is the covariance matrix of the measurement noise $\vec{n}$.
The $i^{th}$ background image $\vec{b}^{(i)}$ was sampled from the probability density function $p(\vec{b}|\vec{g}, H_0)$ by constructing a Markov chain according to the method described in \cite{kupinski2003ideal}.
Each Markov chain was simulated by running 200,000 iterations. 

\vspace{0.8cm}
\subsection{BKS signal detection-localization task with a clustered lumpy background model}
\begin{figure}[H]
\centering
\begin{subfigure}[b]{0.15\textwidth}
   \centering
 \includegraphics[width=1.0\linewidth]{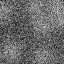}
  \vspace{-0.5cm}
 \caption{}
 \end{subfigure}
 \begin{subfigure}[b]{0.15\textwidth}
  \centering
 \includegraphics[width=1.0\linewidth]{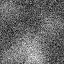}
  \vspace{-0.5cm}
 \caption{}
 \end{subfigure}
  \begin{subfigure}[b]{0.15\textwidth}
  \centering
 \includegraphics[width=1.0\linewidth]{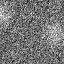}
  \vspace{-0.5cm}
 \caption{}
 \end{subfigure}

\vspace{0.2cm}
  \begin{subfigure}[b]{0.15\textwidth}
   \centering
 \includegraphics[width=1.0\linewidth]{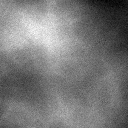}
  \vspace{-0.5cm}
 \caption{}
 \end{subfigure}
 \begin{subfigure}[b]{0.15\textwidth}
  \centering
 \includegraphics[width=1.0\linewidth]{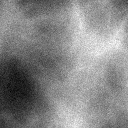}
  \vspace{-0.5cm}
 \caption{}
 \end{subfigure}
  \begin{subfigure}[b]{0.15\textwidth}
  \centering
 \includegraphics[width=1.0\linewidth]{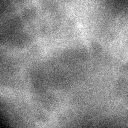}
  \vspace{-0.5cm}
 \caption{}
 \end{subfigure}
 \caption{ (a)-(c) Signal-absent images corresponding to the LB model. (d)-(f) Signal-absent images corresponding to the CLB model.}
 \label{fig:imgs}
 \end{figure}

\begin{figure}[H]
\centering
\begin{subfigure}[b]{0.24\textwidth}
   \centering
 \includegraphics[width=1.0\linewidth]{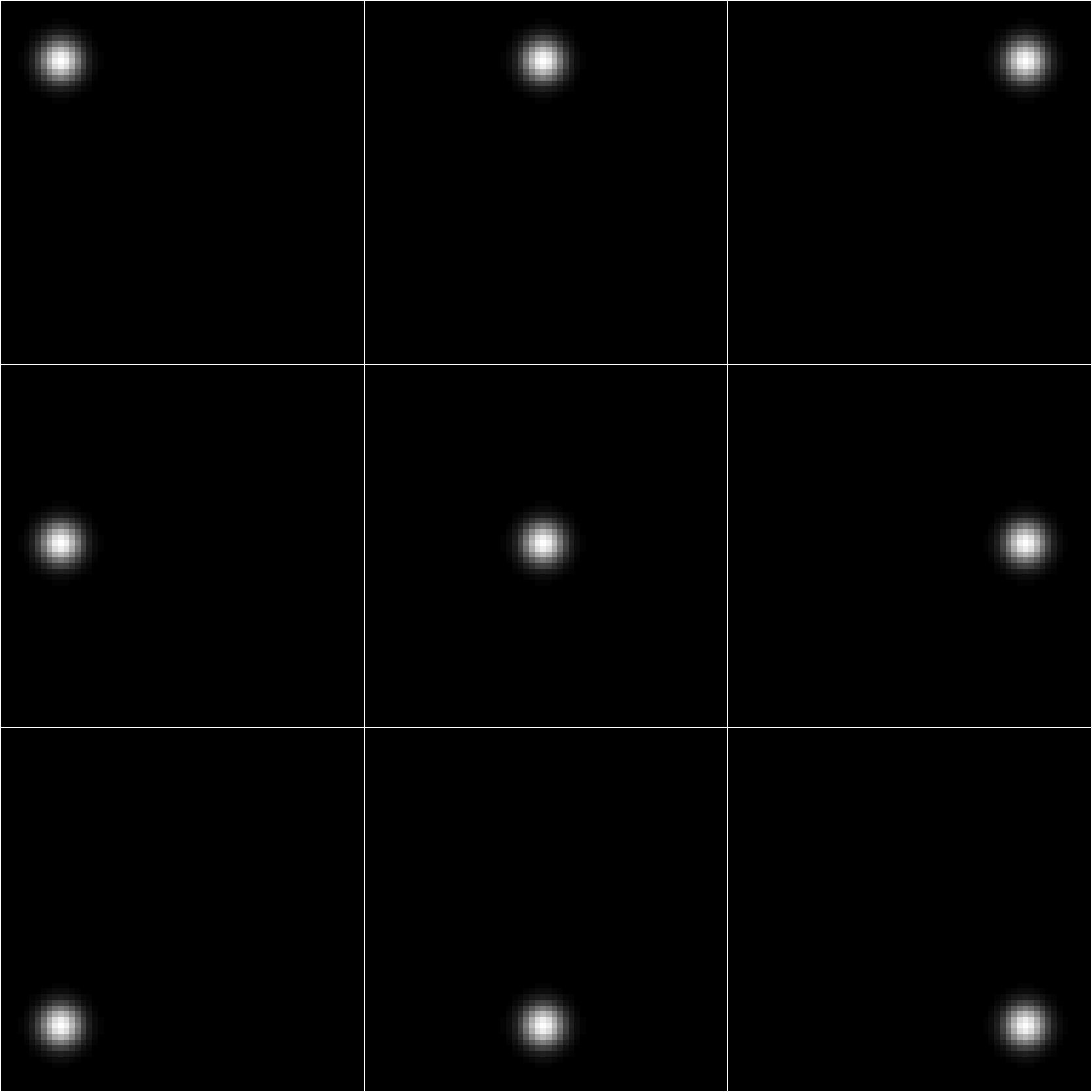}
   \vspace{-0.5cm}
 \caption{}
 \end{subfigure}
 \begin{subfigure}[b]{0.24\textwidth}
  \centering
 \includegraphics[width=1.0\linewidth]{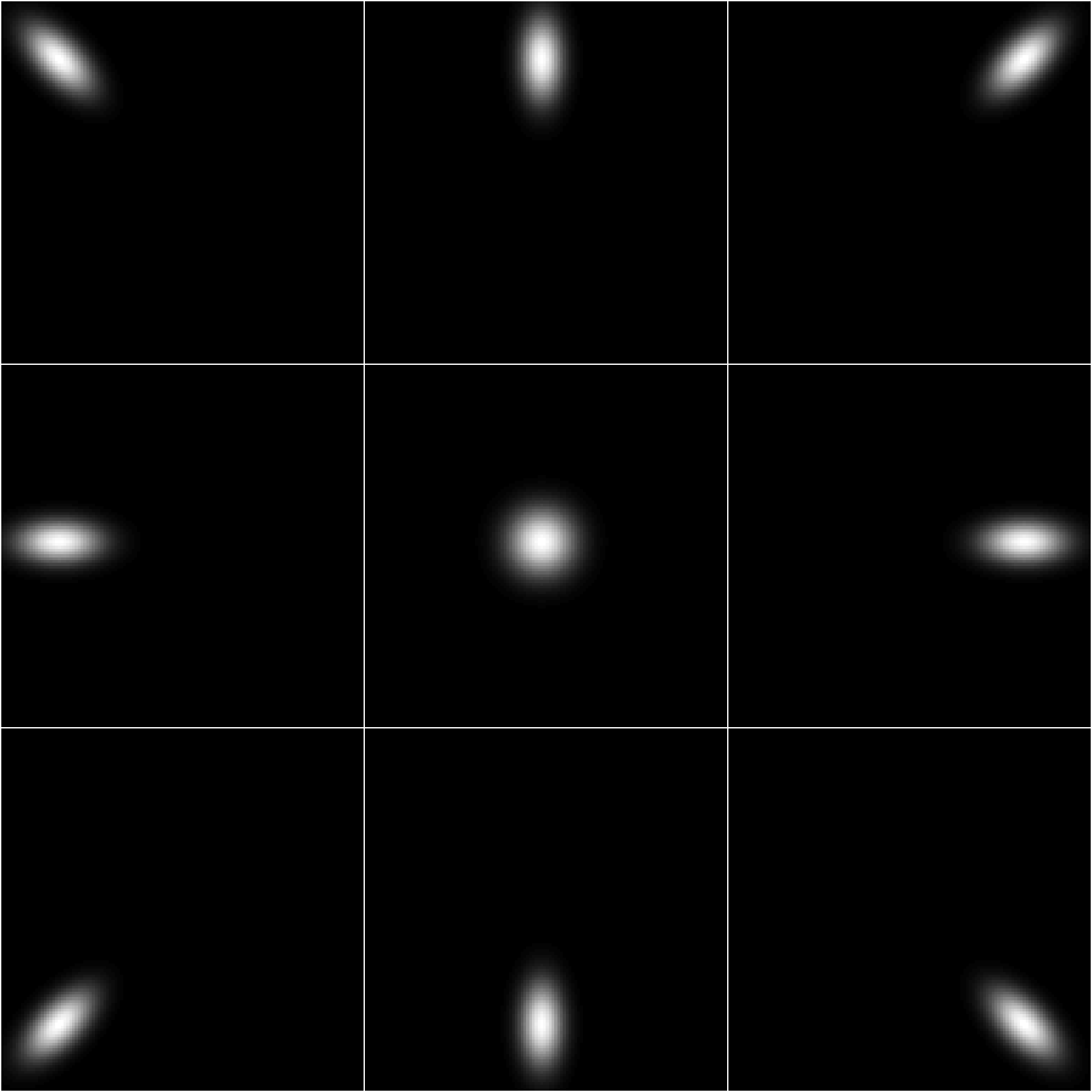}
   \vspace{-0.5cm}
  \caption{}
 \end{subfigure}
  \vspace{-0.5cm}
 \caption{(a) Signal images corresponding to the 9 possible signal locations employed in the BKS task with the LB
model. (b) Signal images corresponding to the 9 possible signal locations employed in the BKS task with the CLB
model.}
 \label{fig:CLB_sigs}
\end{figure}

 The second BKS task utilized a clustered lumpy background (CLB) model to emulate
background variability.  This model was developed  to synthesize mammographic image textures \cite{bochud1999statistical}. 
In this case, the background image $\vec{b}$ had the dimension of $128\times 128$ pixels and its $m^{th}$ element $b_m$ is computed as~\cite{bochud1999statistical}:
\begin{equation}
b_m = \sum_{k=1}^{K}\sum_{n=1}^{N_k} l\left(\vec{r}_m - \vec{r}_k - \vec{r}_{kn}| \mathbf{R}_{\theta_{kn}}\right).
\end{equation} 
Here, $K$ denotes the number of clusters that was sampled from a Poisson distribution with the mean of $\bar{K}$:  $K\sim P(\bar{K})$, $N_k$ denotes the number of blobs in the $k^{th}$ cluster that was sampled from a Poisson distribution with the mean of $\bar{N}$: $N_k\sim P(\bar{N})$,
$\vec{r}_k$ denotes the center location of the $k^{th}$ cluster that was sampled uniformly over 
the image field of view, and
$\vec{r}_{kn}$ denotes the center location of the $n^{th}$ blob in the $k^{th}$ cluster that was sampled from a Gaussian distribution with the center of $\vec{r}_k$ and standard deviation of $\sigma$. 
The blob function $l\left(\vec{r}| \mathbf{R}_{\theta_{kn}}\right)$ was specified as:
\begin{equation}
l\left(\vec{r}|\mathbf{R}_{\theta_{kn}}\right) = A\exp\left(-\alpha \frac{\|\mathbf{R}_{\theta_{kn}} \vec{r}\|^\beta}{L(\mathbf{R}_{\theta_{kn}}\vec{r})} \right),
\end{equation}
where $L(\vec{r})$ is computed as the ``radius" of the ellipse with half-axes $L_x$ and $L_y$, and $\mathbf{R}_{\theta_{kn}}$ is the rotation matrix corresponding to the angle $\theta_{kn}$ that was sampled 
from a uniform distribution between 0 and $2\pi$.
 The parameters of the CLB model employed in this study  are shown in Table. \ref{table1}
\renewcommand{\arraystretch}{1.5}
\begin{table}[H]
\centering
\caption{Parameters for generating CLB images}
\begin{tabular}{| c | c | c | c | c | c | c | c |}
\hline
{$\overline{{K}}$} & $\overline{{N}}$ & $L_x$ & $L_y$ & $\alpha$ & $\beta$ & $\sigma$ & $A$\\ \hline
                       50  & 20  & 5  & 2   & 2.1  & 0.5 & 12 & 40   \\ \hline
\end{tabular}
\label{table1}
\end{table}
The measurement noise was modeled by a mixed Poisson-Gaussian noise model\cite{barrett2013foundations} in which the standard deviation of Gaussian noise was set to 20. Three examples of the signal-absent images are shown in the bottom row in Fig. \ref{fig:imgs}.
 The signal $\vec{s}_j$ had the amplitude of 80,
 the width of $\vec{s}_j$ along each axis took a value from \{5, 8, 10\}, and 
the rotation angle of $\vec{s}_j$ took a value from \{$-\pi/4, 0, \pi/4$\}. 
 The signal at different locations is illustrated in Fig. \ref{fig:CLB_sigs} (b).

\vspace{-0.4cm}
\subsection{CNN training details}
The conventional train-validation-test scheme was employed to evaluate the proposed supervised learning approaches. The CNNs were trained on a training dataset, the CNN architectures and weight parameters were subsequently specified by assessing performance on a validation dataset and,  finally, the performances of the CNNs on the signal detection-localization tasks were evaluated on a testing dataset. The training datasets were comprised of 100,000 lumpy background images and 400,000  CLB background images for the considered BKS detection-localization tasks. Additionally, a ``semi-online learning” method 
in which the measurement noise was generated on-the-fly was employed to mitigate the over-fitting
 problem \cite{zhou2019approximating}. {Specifically, when training CNNs, the training data were simulated by adding measurement noise that was generated on-the-fly to the finite number of noiseless images\cite{zhou2019approximating}. In this way, the number of images employed to train the CNNs could be increased.}
Both the validation dataset and testing dataset comprised 200 images for each class.

\begin{figure}[H]
\centering
\hspace{-0.3cm}{\includegraphics[width=\linewidth]{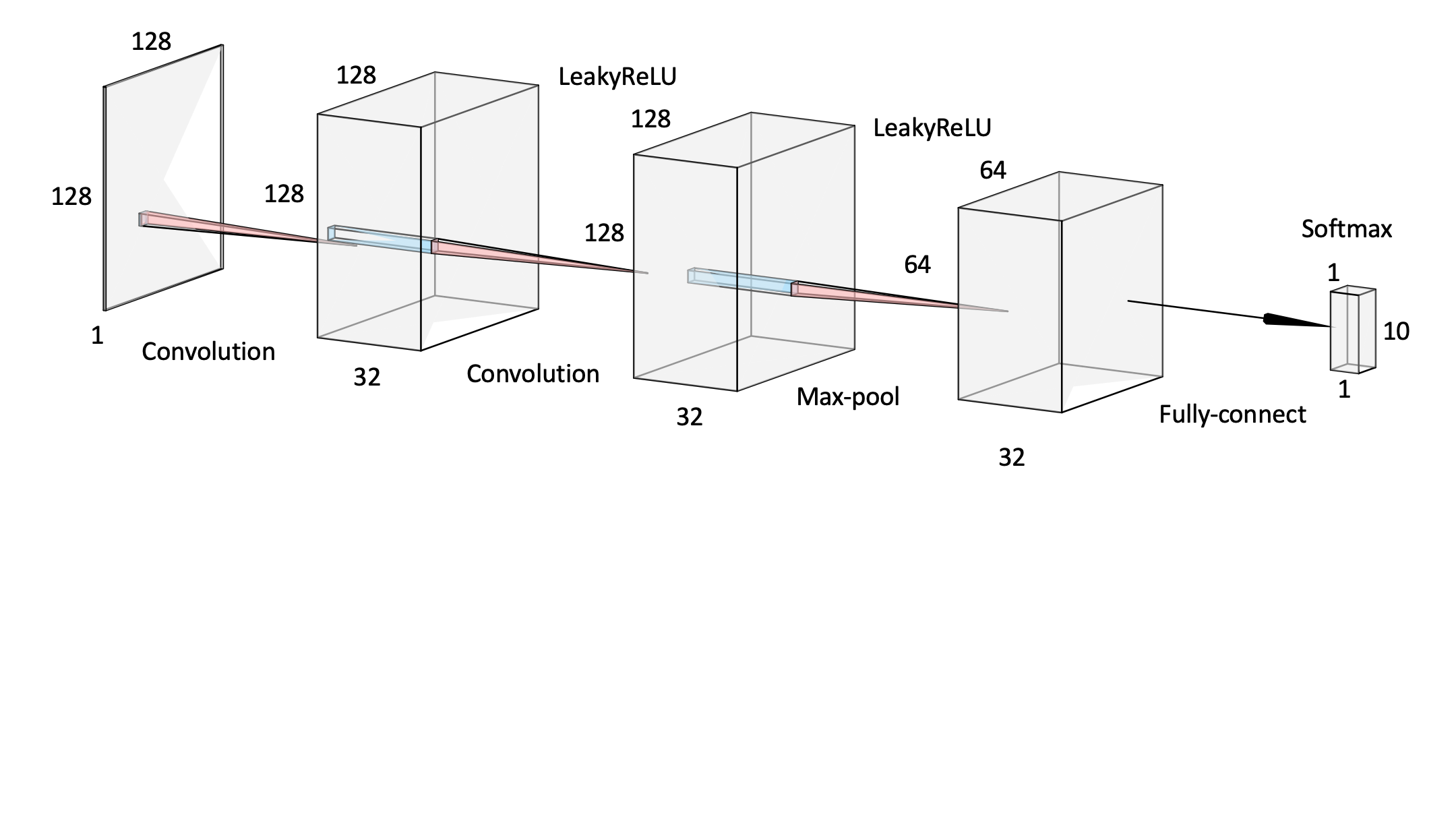}}
\vspace{-2.2cm}
\caption{An instance of the CNN architecture for approximating a set of posterior probabilities for maximizing the ALROC.}
\label{fig:cnn}
\end{figure}

Specifications of CNN architectures that possess the ability to approximate the posterior probability $\Pr(H_j|\vec{g})$ are required.
A family of CNNs that comprise different number of convolutional (CONV) layers was explored to specify the CNN architecture.
Specifically, a CNN having an initial architecture was firstly trained by minimizing the average of the cross-entropy over the training dataset defined Eqn. (\ref{eq:loss}).
CNNs having more CONV layers were subsequently trained until the average of the cross-entropy over the validation dataset did not have significant decrement.
A cross-entropy decrement of at least 1\% of that produced by the previous CNN architecture was considered significant. The CNN that produced the minimum cross-entropy
evaluated on the validation dataset was selected.  All CNN architectures in the considered architecture family comprised CONV layers having 32 filters with the dimension of $5\times 5$, a max-pooling layer\cite{scherer2010evaluation}, and a fully connected layer. A LeakyReLU activation function\cite{springenberg2014striving} was applied to the feature maps produced by each CONV layer
and a softmax function was applied to the output of the fully connected layer.
An instance of the considered CNN architecture is illustrated in Fig. \ref{fig:cnn}. This architecture family was determined heuristically and may not be optimal for many other tasks.
At each iteration of the training, the CNN weight parameters were updated by minimizing the empirical error function on mini-batches 
by use of the Adam algorithm\cite{kingma2014adam}, which is a stochastic gradient-based method.

\section{Results}
\label{sec:result}
\subsection{BKE signal detection-localization task}
Convolutional neural networks that comprised one, three, and five CONV layers were trained for 500,000 mini-batches with each mini-batch comprising 80 images for each class.
For both ``System 1" and ``System 2",
the validation cross-entropy was not significant decreased after 5 CONV layers were employed in the CNNs.
Accordingly,
we stopped training CNNs with more CONV layers, and the CNN corresponding to the smallest validation cross-entropy was selected, which was the CNN having five CONV layer.

\begin{figure}[t!]
\centering
\begin{subfigure}[b]{0.45\textwidth}
   \centering
 \includegraphics[width=1.0\linewidth]{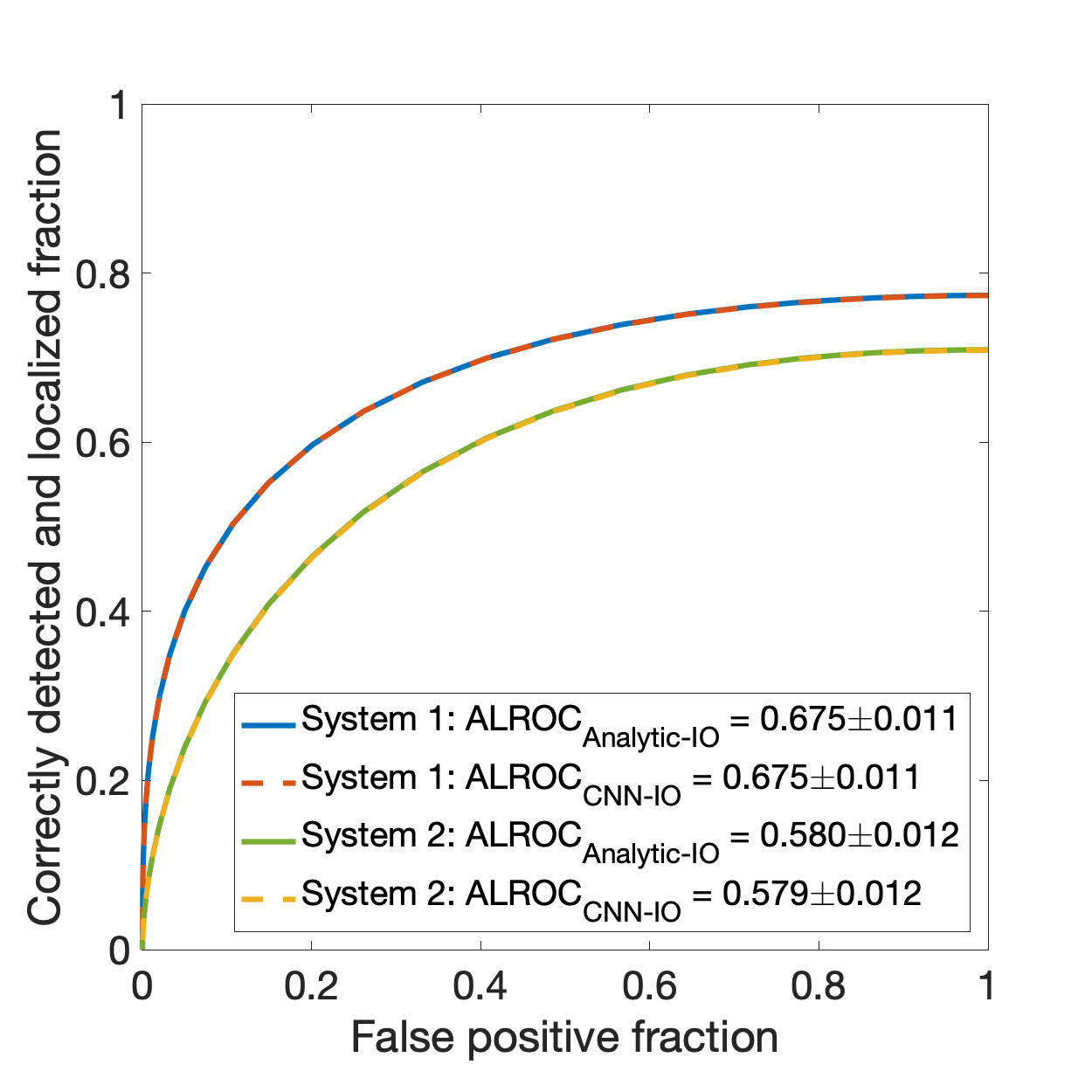}
 \vspace{-0.7cm}
 \caption{}
 \end{subfigure}
 
 \begin{subfigure}[b]{0.45\textwidth}
  \centering
 \includegraphics[width=1.0\linewidth]{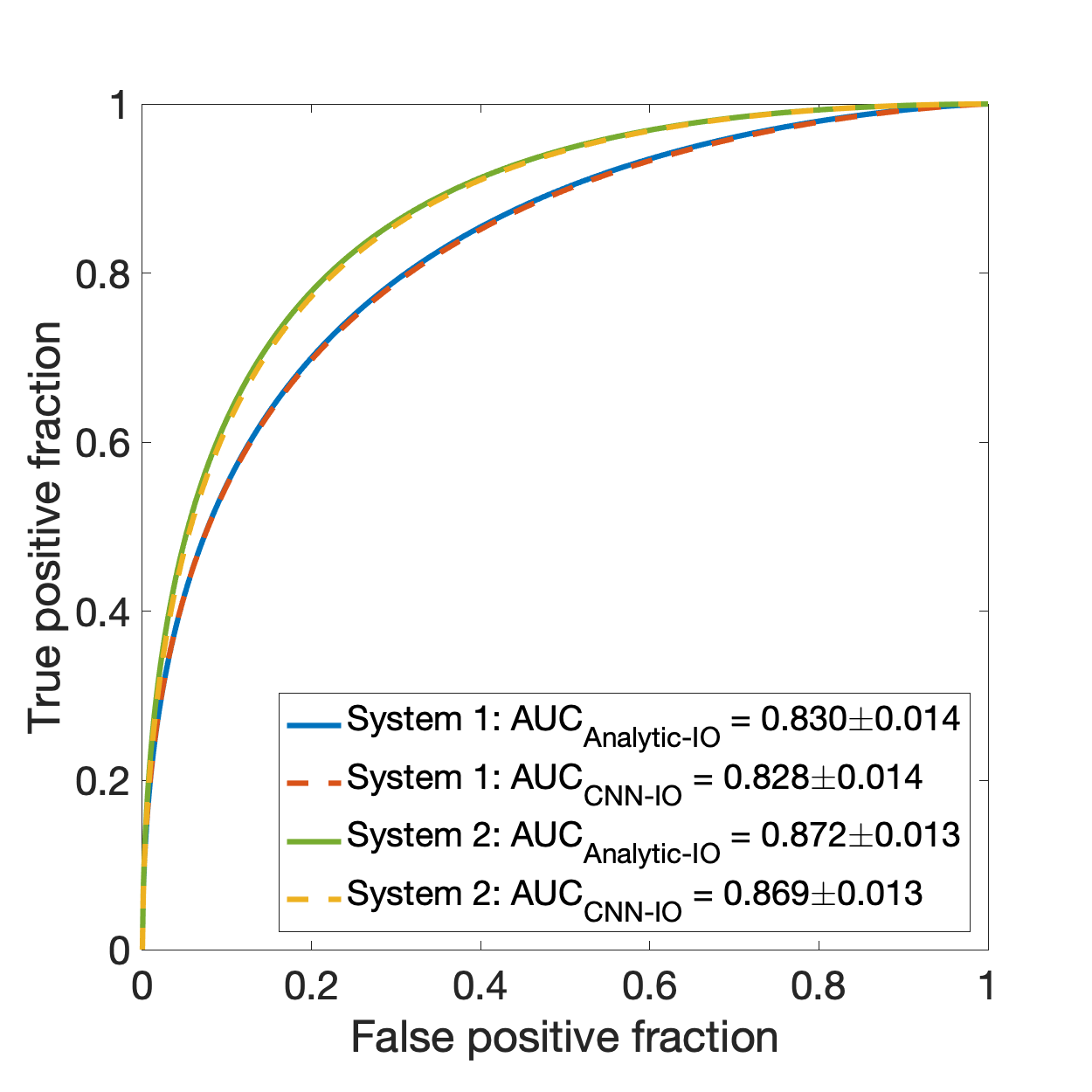}
  \vspace{-0.7cm}
  \caption{}
 \end{subfigure}
 \vspace{-0.2cm} \caption{(a) LROC curves corresponding to the IO for the BKE signal detection-localization tasks. (b) ROC curves corresponding to the IO for  the simplified binary signal detection tasks.}
 \label{fig:BKE}
 \vspace{-0.45cm}
\end{figure}

For the joint detection-localization task, with both imaging systems,
the LROC curves produced by the analytical computation (solid curves) are compared to those produced by the CNN (dashed curves) in Fig. \ref{fig:BKE} (a). 
In addition, for the simplified binary signal detection tasks, the ROC curves  produced by the analytical computation (solid curves) are compared to those produced by the CNN (dashed curves) in Fig. \ref{fig:BKE} (b). 
The curves corresponding to the analytically computed IO and the CNN approximation of the IO (CNN-IO) are in close agreement in both cases.
As shown in Fig. \ref{fig:BKE}, the rankings of the two imaging systems are different when the joint detection-localization task and 
the simplified binary signal detection task were considered.
When the  signal detection-localization task is considered, ``system 1" $>$ ``system 2", while
if the binary signal detection task is considered, ``system 2" $>$ ``system 1".

 \vspace{-0.4cm}
\subsection{BKS signal detection-localization task with a lumpy background model}
Convolutional neural networks comprising 1, 3, 5, 7, 9, and 11 CONV layers were trained for 500,000 mini-batches with each mini-batch comprising 80 images for each class. 
The validation cross-entropy value was not significantly decreased after 11 CONV layers were employed in the CNN, and therefore the CNN having 11 CONV layers 
was selected for approximating the IO.
The performance of the CNN for the signal detection-localization task was characterized by the LROC curve that was evaluated on the testing dataset.
Note that the ALROC value produced by the CNN-IO  was $0.711\pm {0.011}$,
which was larger than the $0.530\pm {0.012}$
 produced by the scanning HO.

\begin{figure}[ht]
  \centering
 \includegraphics[width=0.9\linewidth]{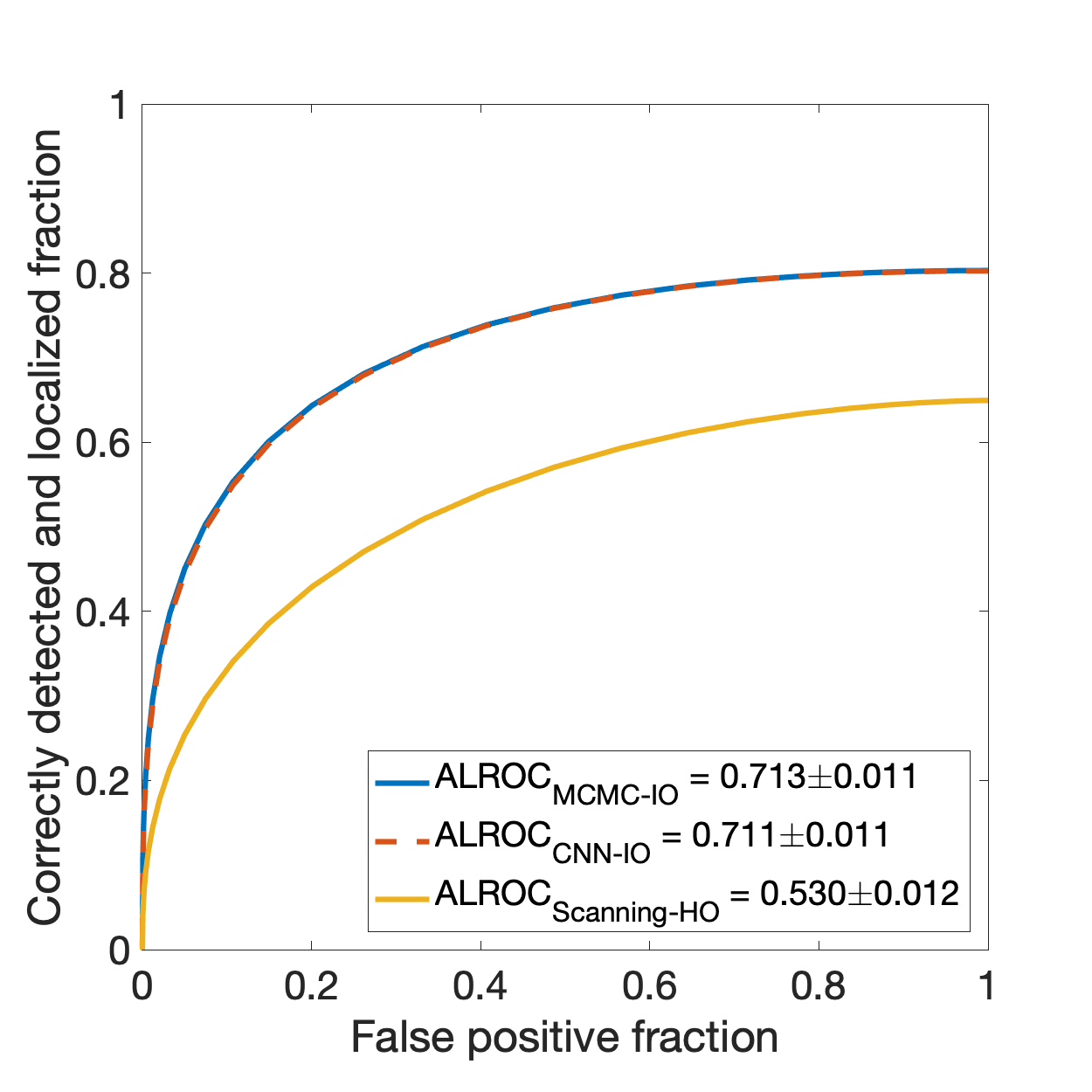}
  \vspace{-0.3cm}
   \caption{The LROC curves produced by the MCMC-IO (blue), CNN-IO (red-dashed), and the scanning HO (yellow) for the BKS task with the lumpy background model. The LROC curve corresponding to the CNN-IO closely approximates that corresponding to the MCMC-IO and is higher than that produced by the scanning HO.}
     \vspace{-0.45cm}
   \label{fig:LB}
\end{figure}

The MCMC simulation provided further validation of the CNN-IO.
 The LROC curve produced by the MCMC method (blue curve) is compared to that produced by the CNN-IO (red-dashed curve) in Fig. \ref{fig:LB}. The curves are in close agreement.
 The ALROC values were $0.713\pm {0.011}$ and $0.711\pm {0.011}$ corresponding to the MCMC and the CNN-IO, respectively.

  \vspace{-0.2cm}
\subsection{BKS signal detection-localization task with a CLB model}
\begin{figure}[H]
  \centering
 \includegraphics[width=0.9\linewidth]{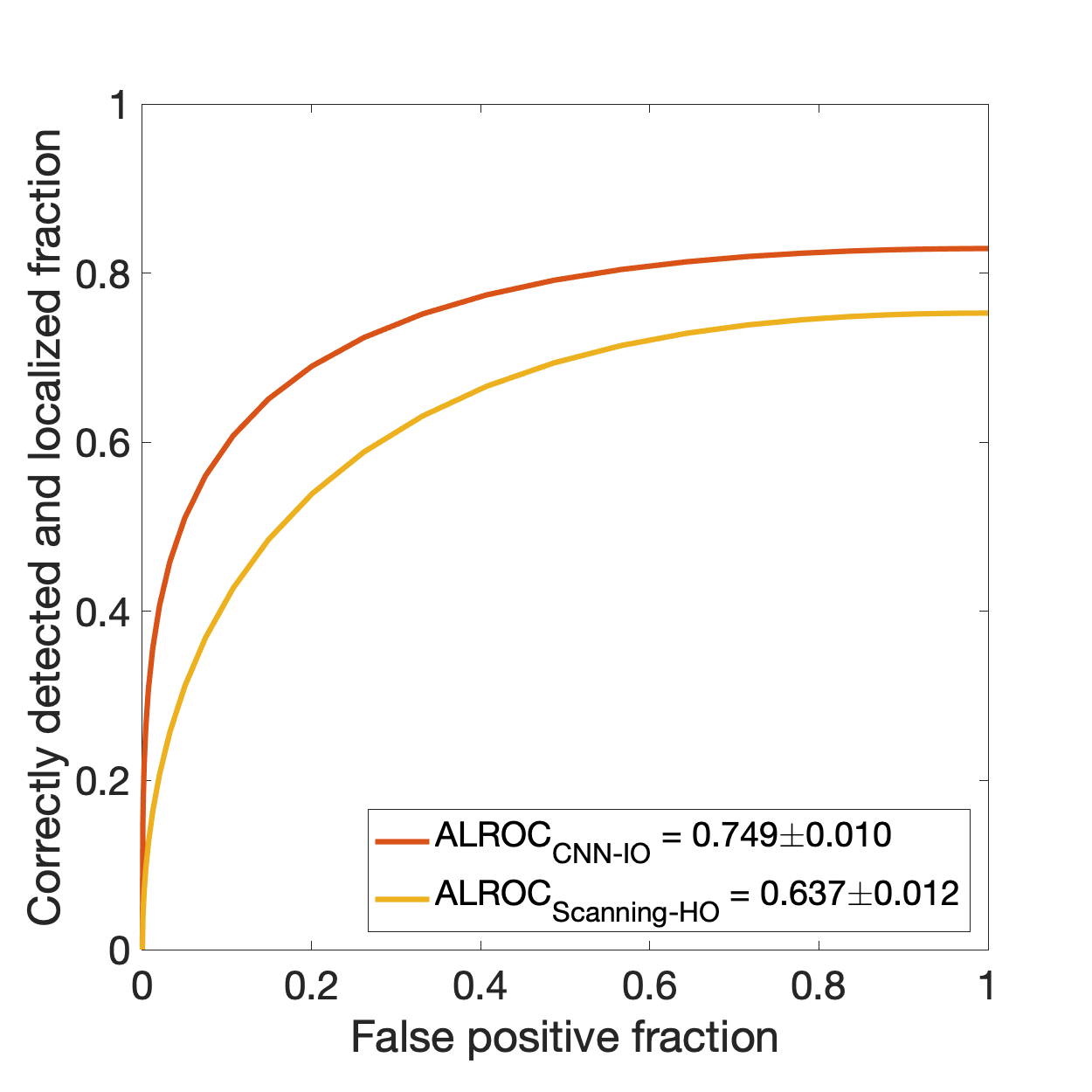}
 \vspace{-0.3cm}
   \caption{The LROC curves produced by CNN-IO (red) and the scanning HO (yellow) for the BKS task with the CLB model. As expected, the LROC curve corresponding to the CNN-IO is higher than that produced by the scanning HO.}
   \label{fig:CLB}
\end{figure}

 \vspace{-0.1cm}
Convolutional neural networks that comprised 1, 3, 5, and 7 CONV layers were trained for 500,000 mini-batches with each mini-batch comprising 20 images for each class. 
The validation cross-entropy value was not significantly decreased after 7 CONV layers were employed in the CNN, and therefore the CNN having 7 CONV layers 
was selected for approximating the IO.
The performance of the selected CNN was quantified by computing the LROC curve and ALROC value on the testing dataset.
The CNN-IO was compared to the scanning HO. The ALROC value produced by the CNN-IO  was $0.749\pm {0.010}$,
which was larger than the $0.637\pm {0.012}$ produced by the scanning HO as expected.
The LROC curves corresponding to the CNN-IO and the scanning HO are displayed in Fig. \ref{fig:CLB}.
\vspace{-0.6cm}
Because the computation of the IO test statistic has not been addressed by MCMC methods for CLB models, 
validation corresponding to MCMC methods was not provided in this case.

\section{Summary}
\label{sec:discussion}
Signal detection-localization tasks are of interest when optimizing medical
imaging systems and scanning numerical observers have been proposed to address them.
However, there remains a scarcity of methods that can be implemented readily for approximating
the IO for detection-localization tasks. 
In this work,
a deep-learning-based  method was investigated to address this need.
Specifically, the proposed method provides a generalized framework for approximating the IO test statistic for multi-class classifications tasks. 
Compared to methods that employ MCMC techniques, supervised learning methods may be easier to implement.
To properly run MCMC methods, numerous practical issues such as the design of proposal densities from which the Markov chain can be efficiently generated need to be addressed.
Because of this, current applications of MCMC methods have been limited to relative simple object models such as a lumpy object model and a binary texture model.
As such, the proposed supervised learning methods may possess a larger domain of applicability for approximating the IO than the MCMC methods.
To demonstrate this, the proposed supervised learning method was applied to approximate the IO for a clustered lumpy object model, 
for which the IO approximation has not been achieved by the current MCMC methods.

{
The proposed supervised learning-based method
may require a large amount of training data to accurately approximate the IO.
Such data may be available when optimizing imaging systems and data acquisition designs
via computer-simulation.
In order to conduct a realistic computer-simulation,
it is desirable to simulate images that capture anatomical variations and textures within a realistic object ensemble.
To achieve this,
one may establish a stochastic object model (SOM) from experimental data by training an AmbientGAN\cite{bora2018ambientgan, zhou2019learning, zhou2020progressively, zhou2020learning}.
Having a well-established SOM, one can produce large amount of training samples to train CNNs to accurately approximate the IO by use of the proposed supervised learning method. Such computer-simulation studies enable the exploration and assessment of a variety of imaging systems and data acquisition designs.
}

{
There remains several other topics for future investigation.
It will be important to quantify the effect of the number of data used in the proposed method on the IO approximation.
In addition, 
to implement the proposed supervised learning methods for approximating the IO in situations where only a limited number of experimental data is available,
it will be important to investigate methods to train deep neural networks on limited training data. To achieve this, one may investigate the methods that employ domain adaptation\cite{ganin2014unsupervised, he2020learning} and transfer learning \cite{qiu2016survey}.
Finally, it will be important to investigate supervised learning methods for approximating IOs for other 
more general tasks such as joint signal detection and estimation tasks associated with the estimation ROC (EROC) curve.
}

\appendices
\section{Gradient of cross-entropy}
\label{append}
The cross-entropy can be written as:
\begin{equation}
\begin{aligned}
& \langle-\log[\Pr(H_y|\vec{g}, \bm\Theta)]\rangle_{(\vec{g}, y)}\\
 = &-\int d\vec{g}\ \left[\sum_{{y}=0}^{J}p(\vec{g}, H_{{y}})\log{\frac{\exp{[z_{{y}}(\vec{g}, \bm\Theta)]}}{\sum_{j' = 0}^{J}\exp{[z_{j'}(\vec{g}, \bm\Theta)]} }}\right]\\
 = &-\int d\vec{g}\ \Big\{\sum_{{y}=0}^{J}p(\vec{g}, H_{{y}}){z_{{y}}(\vec{g}, \bm\Theta)} \\
 &\ \ \ \ \ \ \ - \sum_{{y}=0}^{J}p(\vec{g}, H_{{y}})\log\big(\sum_{{j'}=0}^{J}\exp{[z_{j'}(\vec{g}, \bm\Theta)]} \big) \Big\}.
 \end{aligned}
 \end{equation}
 
{Here, the cross-entropy $\langle-\log[\Pr(H_y|\vec{g}, \bm\Theta)]\rangle_{(\vec{g}, y)}$
is considered as a functional of the $z_j(\vec{g}, \bm\Theta)$, viewed as functions of $\vec{g}$.
The derivative of $\langle-\log[\Pr(H_y|\vec{g}, \bm\Theta)]\rangle_{(\vec{g}, y)}$ with respect to $z_j(\vec{g}, \bm\Theta)$,
which is a functional derivative known as a Fr\'{e}chet derivative,}
 can subsequently be computed as:
 \begin{equation}\label{eq:deriv}
 \begin{aligned}
 &\frac{\partial \langle-\log[\Pr(H_y|\vec{g}, \bm\Theta)]\rangle_{(\vec{g}, y)}}{\partial z_j(\vec{g}, \bm\Theta)}  \\
 &=-p(\vec{g}, H_j) +\sum_{{y}=0}^{J}p(\vec{g}, H_{{y}})\frac{\exp{[z_{j}(\vec{g}, \bm\Theta)]}}{\sum_{j' = 0}^{J}\exp{[z_{j'}(\vec{g}, \bm\Theta)]} } \\
 &=-p(\vec{g}) \left[p(H_j|\vec{g}) - \frac{\exp{[z_j(\vec{g}, \bm\Theta)]}}{\sum_{j' = 0}^{J}\exp{[z_{j'}(\vec{g}, \bm\Theta)]} } \right].
  \end{aligned}
  \end{equation}
  
 The last step in Eqn. (\ref{eq:deriv}) is derived because $p(\vec{g}, H_j) = p(\vec{g}) p(H_j|\vec{g})$ and $\sum_{{y}=0}^{J}p(\vec{g}, H_{{y}}) = p(\vec{g})$.

\bibliography{CNN}{}
\bibliographystyle{IEEETran}

\end{document}